\def\preprintmode{ 1 }	  

\ifnum \preprintmode = 0
   \documentclass[11pt]{article}
\else
   \documentclass[12pt]{article}
\fi

\usepackage{mathptm}
\usepackage[dvips]{graphicx}
\usepackage{color}

\def\insertc#1{                 
 \ifnum \preprintmode = 0
   #1
 \else
 \fi
}

\def\insertp#1{                 
 \ifnum \preprintmode = 0
 \else
   #1
 \fi
}

\begin{document}

\ifnum \preprintmode = 0
 \renewcommand{\baselinestretch}{1} \large\normalsize    
\else
 \renewcommand{\baselinestretch}{1.5} \large\normalsize  
\fi

\pagestyle{myheadings}
\markright{Bedard et al., manuscript of \today}



{

\thispagestyle{empty}       

\renewcommand{\baselinestretch}{1.1} \large\normalsize

\begin{centering}

{\footnotesize
 \begin{tabular}{lr}
   \ \hspace{100mm} \ &
   Manuscript of \today \\
   & Submitted to {\it Biophysical Journal} \\
   & Category: EPS - Electrophysiology 
 \end{tabular}
}

\vspace{25mm}

\ \\ \ \\
{\large\sc
Modeling extracellular field potentials \\
and the frequency-filtering properties
of extracellular space
}

\ \\ \

Claude B\'edard$^1$, Helmut Kr\"{o}ger$^1$ and Alain Destexhe$^{2}$

\ \\ \

1: D\'{e}partement de Physique, Universit\'{e} Laval, \\ Qu\'{e}bec,
Qu\'{e}bec G1K 7P4, Canada

\

2: Unit\'e de Neurosciences Int\'egratives et Computationnelles, CNRS, \\
1 Avenue de la Terrasse, 91198 Gif-sur-Yvette, France

\end{centering}

\

\vspace{25mm}


\insertp{\begin{description}
\item[\hspace{8mm}] Running title:
\begin{quote}
 \it Modeling local field potentials
\end{quote}
\item[\hspace{8mm}] Address for correspondence:
\begin{quote}
 Dr.\ A.\ Destexhe \\
 Unit\'e de Neurosciences Int\'egratives et Computationnelles, \\
 CNRS, \\
 1 Avenue de la Terrasse (Bat.\ 33), \\
 91198 Gif-sur-Yvette, France \\ \ \\
 Tel: 33-1-69-82-34-35, \ Fax: 33-1-69-82-34-27, \\
 email: Destexhe@iaf.cnrs-gif.fr \\
\end{quote}
\end{description}
}

}

\newcommand{\co}[1]{{\bf[#1]}}\newcommand{\be}{\begin{equation}}
\newcommand{\ee}{\end{equation}}
\newcommand{\bea}{\begin{eqnarray}}
\newcommand{\eea}{\end{eqnarray}}
\newcommand{\sgn}{\mbox{sgn}}
\newcommand{\slesssim}{{\scriptstyle \lesssim}}
\renewcommand{\vec}[1]{{\bf #1}}
\newcommand{\sect}[1]{\section*{\large #1}}
\newcommand{\subsect}[1]{\subsection*{#1}}
\newcommand{\subsubsect}[1]{\subsubsection*{\it #1}}



\insertp{\pagebreak}

\sect{Abstract}

Extracellular local field potentials (LFP) are usually modeled as arising from a
set of current sources embedded in a homogeneous extracellular medium.  Although
this formalism can successfully model several properties of LFPs, it does not
account for their frequency-dependent attenuation with distance, a property
essential to correctly model extracellular spikes.  Here we derive expressions
for the extracellular potential that include this frequency-dependent
attenuation.  We first show that, if the extracellular conductivity is
non-homogeneous, there is induction of non-homogeneous charge densities which may
result in a low-pass filter.  We next derive a simplified model consisting of a
punctual (or spherical) current source with spherically-symmetric
conductivity/permittivity gradients around the source.  We analyze the effect of
different radial profiles of conductivity and permittivity on the
frequency-filtering behavior of this model.  We show that this simple model
generally displays low-pass filtering behavior, in which fast electrical events
(such as Na$^+$-mediated action potentials) attenuate very steeply with distance,
while slower (K$^+$-mediated) events propagate over larger distances in
extracellular space, in qualitative agreement with experimental observations. 
This simple model can be used to obtain frequency-dependent extracellular field
potentials without taking into account explicitly the complex folding of
extracellular space.




\insertp{\ \\ \ \\ \ \\ \sect{Keywords}
\begin{quote}
 \it Computational models, Neurons, Local field potentials, LFP,
 Electroencephalogram, EEG
\end{quote}
}



%


\insertc{\pagebreak}
\insertp{\pagebreak}


\sect{INTRODUCTION}
\label{sec:Intro}

Extracellular potentials, such as local field potentials (LFPs) or the
electroencephalogram (EEG), are routinely measured in electrophysiological
experiments.  The fact that action potentials have a limited participation to the
genesis of the EEG or LFPs was noted from early studies.  Bremer (1938, 1949)
proposed that the EEG is generated by non-propagating potentials, based on the
mismatch of time course between EEG waves and action potentials.  Eccles (1951)
proposed that LFP and EEG activities are generated by summated postsynaptic
potentials arising from the synchronized excitation of cortical neurons. 
Intracellular recordings from cortical neurons later demonstrated a close
correspondence between EEG/LFP activity and synaptic potentials (Klee et al.,
1965; Creutzfeldt et al., 1996a, 1996b).  The current view is that EEG and LFPs
are generated by synchronized synaptic currents arising on cortical neurons,
possibly through the formation of dipoles (Nunez, 1981; Niedermeyer and Lopes da
Silva, 1998).

The fact that action potentials do not participate to EEG-related activities
indicate strong frequency-filtering properties of cortical tissue.  High
frequencies (greater than $\approx$~100~Hz), such as that produced by action
potentials, are subject to a severe attenuation, and therefore are visible only
for electrodes immediately adjacent to the recorded cell.  On the other hand,
low-frequency events, such as synaptic potentials, attenuate less with distance. 
These events can therefore propagate over large distances in extracellular space
and be recordable as far as on the surface of the scalp, where they can
participate in the genesis of the EEG.  This frequency-dependent behavior is also
seen routinely in extracellular unit recordings: the amplitude of
extracellularly-recorded spikes is very sensitive to the position of the
electrode, but slow events show much less sensitivity to the position.  In other
words, an extracellular electrode records slow events that originate from a large
number of neighboring neurons, while the action potentials are recorded only for
the cell(s) immediately adjacent to the electrode.  This fundamental property
allows to resolve single units from extracellular recordings.

However, little is known about the physical basis of the frequency-dependent
attenuation of extracellular potentials in cortex.  By contrast to intracellular
events, which biophysical mechanisms have been remarkably well characterized
during the last 50 years (reviewed in Koch, 1999), comparatively little has been
done to investigate the biophysical mechanisms underlying the genesis of
extracellular field potentials (see review by Nunez, 1981).  The reason is that
LFPs result from complex interactions involving many factors, such as the spatial
distribution of current sources, the spatial distribution of positive and
negative electric charges (forming dipoles), their time evolution (dynamics), as
well as the conductive and permittivity properties of the extracellular medium. 
One of the simplest and widely used model of LFP activity considers current
sources embedded in a homogeneous extracellular medium (Nunez, 1981; Koch and
Segev, 1998).  Although this formalism has been successful in many instances
(Rall and Shepherd, 1968; Klee and Rall, 1977; Protopapas et al., 1998; Destexhe,
1998), it does not account for the frequency-dependent attenuation and therefore
is inadequate for modeling extracellular field potentials including spike
activity.

In this paper, we would like to investigate possible physical grounds for the
frequency-filtering properties of LFPs.  We start from first principles (Maxwell
equations) and consider different conditions of current sources and extracellular
media.  We delineate the cases leading to frequency-filtering properties
consistent with physiological data.  We show that the assumption of a resistive
homogeneous extracellular medium cannot account for the frequency-dependent
attenuation.  It is necessary to take into account the inhomogeneous structure of
the extracellular medium (in both permittivity and conductivity) in order to
account for frequency-dependent attenuation.  We next analyze a simplified
representation of current sources in non-homogeneous media, and provide a
simplified model which could be applied to simulate extracellular field
potentials without using complex representations of extracellular space.  We
terminate by showing a concrete example of the genesis of extracellular LFPs from
a conductance-based spiking neuron model.



\sect{MATERIAL AND METHODS}
\label{sec:Material}

We will first develop a general formalism to express the temporal variations of
extracellular potential, as well as a simple model in which most of the
calculations can be done analytically (see Section {\it General theory}; see also
Appendix~1 and 2 for details).  We will next explore this simplified model
numerically to illustrate its frequency-filtering behavior (Section {\it
Numerical simulations}), in which we have performed two computations: (i)
Calculate the impedance: the impedance is given by an integral, which was
evaluated numerically by standard numerical integration routines. (ii) Convert
time-dependent functions into frequency-spectra.  These conversions were done via
Fourier transformation (as well as its reverse transformation), which were
carried out in C using standard numerical routines (Press et al., 1986). 

To test this formalism, we also considered a simple biophysical model of a
spiking neuron containing voltage-dependent and synaptic conductances (last part
of Section {\it Numerical simulations}).  A single-compartment neuron was
constructed and included conductance-based models of voltage-dependent
conductances and synaptic conductances.  This model was described by the
following membrane equation:
\be
  C_m {dV \over dt} \ = \ -g_L (V-E_L) \ -g_{Na} (V-E_{Na}) \ -g_{Kd} (V-E_K) \
                          -g_{M} (V-E_K) \ -g_e (V-E_e) ~ ,
\ee
where $C_m$ = 1~$\mu$F/cm$^2$ is the specific membrane capacitance, $g_L$ = $4.52
\times 10^{-5}$~S/cm$^2$ and $E_L$ = -70~mV are the leak conductance and reversal
potential.  $g_{Na}$ = 0.05~S/cm$^2$ and $g_{Kd}$ = 0.01~S/cm$^2$ are the
voltage-dependent Na$^+$ and $K^+$ conductances responsible for action potentials
and were described by a modified version of the Hodgkin \& Huxley (1952) model. 
$g_M$ = $5 \times 10^{-4}$~S/cm$^2$ is a slow voltage-dependent K$^+$ conductance
responsible for spike-frequency adaptation.  $g_e$ = 0.4~$\mu$S is a fast
glutamatergic (excitatory) synaptic conductance.  The voltage-dependent
conductances were described by conventional Hodgkin-Huxley type models adapted
for modeling neocortical neurons, and the synaptic conductance was described by a
first-order kinetic model of neurotransmitter binding to postsynaptic receptors. 
These models and their kinetic parameters were described in detail in a previous
publication (Destexhe and Par\'e, 1999).  All numerical simulations were
performed using the NEURON simulation environment (Hines and Carnevale, 1997).



\sect{GENERAL THEORY}
\label{sec:Theory}

In this section, we outline the main features of the model starting from first
principles (Maxwell equations).  We will consider a number of different special
cases and derive a simplified model with radial (spherical) symmetry.  In Section
{\it Numerical simulations}, we will investigate numerically the behavior of this
simplified model.

In Maxwell's theory, the electric properties of a conductive medium are
determined by two parameters, conductivity $\sigma$ and permittivity $\epsilon$. 
While conductivity quantifies the local relation between the electric field and
the current, permittivity characterizes the response of the system in terms of
separation of opposite charges (polarisation) in the presence of an electric
field. Maxwell's theory of electromagnetism allows one to compute electric and
magnetic fields or potentials, for a given distribution of charges and electric
currents. Because charges move very slowly in biological media, the effects of
magnetic fields are very small compared to that of the electric field and
will be neglected here.

One of Maxwell's equations is {\it Gauss' law}:
\be
\label{eq:Gauss}
\vec{\nabla} \cdot (\epsilon ~ \vec{E})
= \rho  ~ .
\ee
Here $\vec{E}$ denotes is the electric field, $\vec{D} = \epsilon \vec{E}$ is
the called displacement and $\rho$ is the charge density of the extracellular
medium, also allowed to vary slowly in time.

The {\it continuity equation} relates the current density $\vec{j}$ to the
charge density $\rho$:
\be \label{eq:Continuity}
 \vec{\nabla} \cdot\vec{j}+\frac{\partial \rho}{\partial t}=0
  \ee
which states a balance between the electric flux into some volume and the
change of the total charge in this volume. In other words, no charge will get
lost.

Finally, there is {\it Ohm's law}:
\be
\label{eq:Ohm}
\vec{j} = \sigma ~ \vec{E} ~ ,
\ee
which states that the relation between the local electric field and the local
current density can be described by a single macroscopic parameter $\sigma$. 
Here, we assume that the electric field and the current are parallel locally
(i.e., at any given point in the medium). The coefficient relating $\vec{E}$ and
$\vec{j}$, the conductivity $\sigma$, is a scalar, which is justified by
macroscopic measurements (Ranck, 1963).

Combining Eqs.~\ref{eq:Continuity} and \ref{eq:Ohm} yields
\be \label{eq:ChargeConserv} \vec{\nabla} \cdot (\sigma ~ \vec{E})
+ \frac{\partial \rho}{\partial t}=0
 ~ . \ee
This equation describes charge conservation in
differential form.

In the following, we assume that electric currents are distributed on the
surface of the membrane (ionic currents), and that these currents are
allowed to vary in time.  We introduce the the electric potential $V$
(also called extracellular potential) which obeys
\be
\label{eq:Pot}
\vec{E} = - \vec{\nabla} V ~ .
\ee

Below, we successively consider different cases of increasing complexity,
starting with a homogeneous extracellular medium, then going over to
non-homogeneous media.

\subsect{Homogeneous extracellular medium}
\label{sec:HomogMed}

Consider a membrane embedded in a homogeneous extracellular
medium, with conductivity $\sigma$ and permittivity $\epsilon$
being held constant in space and time. As shown in Appendix~3, using the 
assumption $\sigma = \sigma_{0} = const.$, $\epsilon = \epsilon_{0} = const.$, 
we get for each spectral component
\be \label{eq:CompPot} (\vec{\nabla} V_{\omega}) \cdot
(\frac{\vec{\nabla} (\sigma + i \omega \epsilon)}{(\sigma + i
\omega \epsilon)}) + \Delta V_{\omega} =\Delta V_{\omega}= 0 ~ .
\ee
Hence we find
\be \label{eq:Poisson} \Delta V_\omega = -\frac{\rho_\omega}{\epsilon}=0 ~ . \ee
Using the inverse Fourier transform, this yields
\be \label{eq:Laplace} \Delta V(\vec{x},t) = -\frac{\rho}{\epsilon}= 0 ~ . 
\ee
One observes that the charge density $\rho$ vanishes at the exterior of the
sources.  The solution depends on the geometry considered, its symmetries and
boundary conditions.  We consider different cases below.

\subsubsect{Spherical membranes}
\label{sec:SpherMembr}

As a particular case, let us consider a spherical membrane obeying the
conditions outlined above (homogeneous medium of constant conductivity and
permittivity), in which the potential is allowed to vary as a function of
time.  There are two classic cases of boundary conditions, which yield unique
solutions of the Laplace equation: (i) Specification of the potential on a
surface (Dirichlet boundary condition) and (ii) Specification of the
derivative of the potential normal to the surface (Neumann boundary
condition). The latter is equivalent to specifying the normal component of the
electric field. Due to Ohm's law this means to specify the current density on
the surface.  In neurons, we consider the Neumann conditions in which current
sources represent the ionic currents in the membrane.  In this case, we have
the following boundary conditions:
\be
\label{eq:NeumannBound}
\vec{j}(r=R) = j_{0} \vec{n} = \sigma \vec{E}(r=R) ~ ,
\ee
Here, we have assumed that the membrane is spherical of radius $R$ (which
is also equivalent to consider a point current source).  The current density
on the surface of the sphere is is proportional to the unit vector $\vec{n}$
normal to the surface of the sphere.

To analyze the behavior of the extracellular potential as a function of
frequency, we perform a Fourier transformation with respect to time of the
potential,
\be \label{eq:Fourier} V_{\omega}(\vec{x} ) =
\int_{-\infty}^{\infty} dt ~V(\vec{x},t) e^{i \omega t} ~ . \ee
The potential $V$ satisfies the Laplace equation (11). Because the
Fourier transform is linear, each Fourier component $V_{\omega}$
also satisfies (10). The solution, satisfying boundary conditions
(Eq.~\ref{eq:NeumannBound}), is given by
\be \label{eq:SpherSol} V_{\omega}(r) = \frac{I_\omega}{4 \pi
\sigma r} ~ , \ee
where $I_\omega = j_{\omega 0}/(4 \pi R^{2})$ is the total current for each
frequency component. This shows that the potential is the same for all frequency
components.  Therefore, there is no frequency-dependence in this case.

Thus, the general expression for field potentials resulting from a set of $N$ 
current sources $\{I_j\}$ of spherical symmetry is:
\be
V(\vec{x}) \ = \ {1 \over 4 \pi \sigma} \ \sum_{j=1}^N \
\frac{I_j}{|\vec{x}-\vec{x}_j|} ,
\ee
where $\vec{x}$ is a point in extracellular space, $\vec{x}_j$ is the location
of the $j$th current source, and $|\vec{x}-\vec{x}_j|$ is the distance between
$\vec{x}$ and $\vec{x}_j$. This expression is widely used to model
extracellular field potentials (Nunez, 1981; Koch and Segev, 1998).

\subsubsect{Cylindric membranes}

Because neuronal processes (dendrites, axons) are closer to cylinders, we
considered cylindric membranes as a second particular case.  Here, the
procedure is similar to the above, but the geometry and symmetries are different.
If we assume that the membrane potential respects cylindric symmetry (ie, $V$
does not depend on the rotation angle around the cylinder axis), and is uniform
on the surface of the membrane (isopotential compartment), then
Eq.~\ref{eq:Laplace} can be solved for each component of the frequency
spectrum and has a unique solution:
\be
\label{eq:CylSol}
V_{\omega}(r) \ = \ V_{\omega}(R) \ + \ j_\omega(R) \ \frac{R}{\sigma} \ \ln{(r/R)} ~ ,
\ee
where $r$ is the distance perpendicular to the cylinder axis, while $j_\omega(R)$
and $V_{\omega}(R)$ are the $\omega$-frequency components of the current density
and potential at the surface of the cylinder.

In this case again, the extracellular potential is independent of the frequency.
The same conclusion applies to membranes of arbitrary geometries taken in the
same conditions because the $\omega$-frequency component of the current density
and potential obey Laplace equation.  There is therefore no frequency dependence
arising from single current sources in homogeneous media.

\subsect{Non-homogeneous extracellular medium}
\label{sec:NonHom}

We have shown above that, in a homogeneous medium with a current source of
spherical or cylindric symmetry, the extracellular electric potential is the
same for all frequency components, and therefore cannot display
frequency-dependent properties.  We now turn to a possible source of
frequency-dependent attenuation, namely the presence of inhomogeneities in
the conductivity of the extracellular medium.

\subsubsect{Stationary currents in spherically-symmetric non-homogeneous medium}
\label{sec:StatCurrNonHom}

Before investigating the general case, let us first consider the case of a static
spherical current source embedded in a medium where the conductivity $\sigma$
conserves spherical symmetry, but varies as a function of distance $r$ (as above
we assume that $\sigma$ does not depend on time).  We also continue to assume
that permittivity $\epsilon$ is homogeneous.  If the total current flowing through
the sphere of radius $R$ is denoted by $I$, then the radial dependence of the
current density is given by
\be
\label{eq:Current}
\vec{j}(r) = \frac{I}{4 \pi r^{2}} ~ \vec{e}_{r} ~ .
\ee
In this case, the charge density $\rho$ is non-zero and is given by
\be
\rho = - \frac{\epsilon}{\sigma} ~ \vec{j} \cdot \vec{\nabla} \log \sigma ~ .
\ee
Then Ohm's law implies for the spherically symmetric electric field
\be
\label{eq:EField}
\vec{E}(r) =  \frac{I}{4 \pi r^{2} \sigma(r) } ~ \vec{e}_{r} ~ .
\ee
The spherically symmetric electric potential is the obtained by integrating
the electric field, giving
\be V(r) = -\int_{\infty}^{r} dr' ~ E(r') =  \int_{r}^{\infty} dr'
~ \frac{I}{4 \pi r'^{2} \sigma(r')} ~ . \ee
Details of the calculation can be found in Appendix~1.

This equation shows that the potential may decrease or even increase, depending
on the spatial variations of $\sigma$. An important consequence is that such net
charge creates its own electric field (so-called secondary field), which will be
analyzed in more detail below.  What is the physical origin of this non-zero net
charge? The current density behaves like ${\it J} \propto 1/r^{2}$,
(Eq.~\ref{eq:Current}), and the electric field like ${\it E} \propto 1/(\sigma(r)
r^{2})$, (Eq.~\ref{eq:EField}). Consequently, there will be accumulations of
charges in some regions of lower conductivity, similar to traffic jams.  Consider
a more realistic case in which the conductivity of the extracellular space is
constant on average, but displays spatial fluctuations around this average. This
could correspond for example to different processes and obstacles in the
extracellular medium. In this case, the electric field , going like $1/r^{2}$ on
average, fluctuates locally. This creates local areas of positive and negative
charge, i.e.\ electric dipoles. Those dipoles also create a secondary electric
field.  However, to account for frequency dependence, time-varying current
sources must necessarily be considered, in which case the situation is more
complex, as analyzed in the next section.

\subsubsect{Time-varying currents in non-homogeneous medium}
\label{sec:TimeDepCurr}

Let us now consider the general case where both $\epsilon$ and $\sigma$ are
non-homogeneous in space, but constant in time.  We assume that the current
source is allowed to vary in time.  The continuity equation implies that the
charge density is also time-dependent.  Ohm's law implies that the electric
field has a time-dependence as well, and so will also the extracellular
potential.  Due to the inhomogeneity of $\epsilon$, the extracellular potential
does no longer satisfy Poisson's equation.  To study the frequency-dependence
of the extracellular potential, we perform a Fourier transform of the electric
field, the potential and likewise of the charge density $\rho$.

$\rho_{\omega}$, the component of frequency $\omega$ of the temporal
Fourier transform of the charge density $\rho$, satisfies
\be
\label{eq:FourierCharge}
\frac{\partial }{\partial t} \rho_{\omega} = i \omega \rho_{\omega} ~ .
\ee
This equation expresses the differential law of charge conservation for a
given Fourier component.  Now we consider the Gauss' law (Eq.~\ref{eq:Gauss}), the
law of charge conservation in differential form (Eq.~\ref{eq:ChargeConserv}), and
carry out the Fourier transform with respect to time. Taking into account
Eq.~\ref{eq:FourierCharge} yields
\be
\label{eq:FourPot}
\Delta V_{\omega} = - \frac{
(\vec{\nabla} V_{\omega}) \cdot (\vec{\nabla} (\sigma + i \omega \epsilon)) }
{ \sigma + i \omega \epsilon }
= - (\vec{\nabla} V_{\omega}) \cdot
(\vec{\nabla} \log(\sigma + i \omega \epsilon)) ~ .
\ee
For details, see Appendix~2.

This equation is general and applies to any particular symmetries (under the
assumption of scalar conductivity).  We consider below a series of special cases,
as well as special symmetries.

\subsubsect{Special cases}
\label{sec:cases}

As a first special case, consider Eq~\ref{eq:FourPot} when permittivity is
constant. Then Ohm's law (Eq.~\ref{eq:Ohm}) implies
\be
\Delta V_{\omega} = - \frac{\rho_\omega}{\epsilon} ~ .
\ee
Constant permittivity also implies
$\nabla(\sigma + i \omega \epsilon) = \nabla(\sigma)$.
Hence Eq.~\ref{eq:FourPot} takes the form
\be
\label{eq:ConstPerm}
\Delta V_{\omega} = - \frac{ \vec{\nabla} V_{\omega} \cdot \vec{\nabla} \sigma }
{ \sigma + i \omega \epsilon } = - \frac{\rho_\omega}{\epsilon} ~ .
\ee
We therefore observe the occurrence of a (complex) phase difference between
the induced charge density $\rho/\epsilon$ and the current density $j$
(recall: $- \vec{\nabla} V_{\omega} = \vec{E}_{\omega} =
\vec{j}_{\omega}/\sigma$).  This effect depends on the frequency $\omega$ of
the Fourier component.  Such phenomenon is well known from electric circuits
of the RC type, where in general a phase difference between potential and
current is observed. In particular, if the potential vanishes at some time
$t$, the electric charge density will not immediately go to zero.

Eq.~\ref{eq:ConstPerm} shows that for high enough frequency the induced charge
density goes to zero.  On the other hand, for low-frequency phenomena, the charge
density will carry out large fluctuations and will be sensitive to spatial
fluctuations of conductivity.  This has important consequences for interpreting
LFP activity (see Section {\it Frequency-filtering properties of non-homogeneous
media}).

As another special case, consider Eq.~\ref{eq:FourPot} when the conductivity
is constant. The law of charge conservation in differential form
Eq.~\ref{eq:ChargeConserv} then becomes
\be \Delta V_{\omega} =  \frac{1}{\sigma} \frac{\partial
\rho_\omega}{\partial t} ~ . \ee
Constant conductivity also implies
$\nabla(\sigma + i \omega \epsilon) = \nabla(i \omega \epsilon)$.
Hence Eq.~\ref{eq:FourPot} takes the form
\be
\label{eq:ConstCond}
\Delta V_{\omega} =
- \frac{ \vec{\nabla} V_{\omega} \cdot \vec{\nabla} i \omega \epsilon }
{ \sigma + i \omega \epsilon } =
 \frac{1}{\sigma} \frac{\partial \rho_\omega}{\partial t}
= - \frac{1}{\sigma} \vec{\nabla} \cdot \vec{j} ~ .
\ee
This means in the limit of low frequencies that there are no current sinks or
sources.

A third noteworthy special case is when both permittivity and conductivity are
non-homogeneous, but have a fixed ratio:
\be
\frac{\epsilon}{\sigma} = const.
\ee
Under those circumstances one obtains
\be
\label{eq:NoFourPot}
\Delta V_{\omega} =
- \frac{ (\vec{\nabla} V_{\omega}) \cdot (\vec{\nabla} \sigma) }
{ \sigma  }
= - (\vec{\nabla} V_{\omega}) \cdot
(\vec{\nabla} \log(\sigma)) ~ .
\ee
All frequency-dependence cancels out, the potential becomes frequency
independent.

\subsubsect{Frequency-filtering properties of non-homogeneous media}
\label{sec:filter}

We now analyze the frequency-filtering properties of Eq.~\ref{eq:FourPot}. 
Consider this equation in two limit cases:
(i) $\omega << \sigma/\epsilon$ (low frequency limit); in this case,
$\log(\sigma + i \omega \epsilon) \approx \log(\sigma)$, the solution becomes
independent of the permittivity $\epsilon$ and is determined only by the
conductivity $\sigma$.  (ii) $\omega >> \sigma/\epsilon$ (high frequency
limit); in this case, $\log(\sigma + i \omega \epsilon) \approx \log(i \omega
\epsilon)$, and permittivity only determines the solution. These two cases
will be considered in more detail below.  The critical frequency around which
this transition will occur depends on the relative values of $\sigma$ and
$\epsilon$, and one can define the following critical frequency $f_{cr}$
\be
f_{cr} = 2\pi \omega_{cr}, ~~~ \omega_{cr} = \frac{\sigma}{\epsilon} ~ .
\ee
As an example, consider the value of average resistivity $\rho_{res}$ (inverse of
conductivity $\sigma$) measured in rabbit cerebral cortex (Ranck, 1963), giving
$\rho_{res} = 3\Omega~m$.  Taking the permittivity of salt water ($\epsilon = 7
\times 10^{-10} F/m$), gives a critical frequency of $f_{cr}$ of about
$10^{10}$~Hz. Thus, this analysis shows that the behavior will be similar for low
and high frequency limits. However, if one evaluates ${2\pi\sigma}/{\epsilon}$
for a resting membrane (closed ion channels; $\rho\simeq 10^{9}\Omega m$ and
$\epsilon\simeq 10^{-10} F/m$), one finds for the critical frequency a value in
the range between $0$ and $100$ Hz.  The phenomenon of induced charges will be
likely to play a role in the frequency range of synaptic inputs in cerebral
cortex (0 to 40~Hz). On the other hand, higher frequencies ($>100~Hz$) -- such as
action potentials -- are likely to cause negligible variations in charge density.

Further, we can compare Eq.~\ref{eq:FourPot} to cases where there is no frequency
dependence (i.e., Laplace Eq.~\ref{eq:Laplace}).  At the limit of high
frequencies, the left term of Eq.~\ref{eq:FourPot} vanishes and this equation
becomes equivalent to Eq.~\ref{eq:Laplace}, showing that for high frequencies,
one recovers the same behavior as for a homogeneous medium.  To have a low-pass
filter, similar to what is observed from extracellular recordings, one must have
a situation in which the attenuation of the potential at low frequencies must be
{\it less} than for homogeneous media.  Inspection of the left term of
Eq.~\ref{eq:FourPot} shows that the attenuation can be either less or more
pronounced, resulting in low- or high-pass filters.  The type of filter will
depend on the behavior of the gradients of conductivity and permittivity.  This
behavior will be analyzed numerically in more detail later (see Section {\it
Numerical simulations}).

\subsubsect{Time-varying currents in spherically-symmetric non-homogeneous medium}
\label{sec:SpherNonhom}

In order to calculate the extracellular potential $V$ generated by time-varying
currents in non-homo\-ge\-neous media, Eq.~\ref{eq:FourPot} must be integrated by
incorporating details about the particular geometry of current sources and
extracellular properties. A general method for solving this problem is, e.g.,
finite-element analysis, which allows to explicitly incorporate the complex shape
and composition of extracellular space around neurons. However, this approach
requires to integrate complex morphological data and appropriate simulation
tools. We defer this to a future study.  In order to have a model of LFPs
applicable to standard neuron models, we follow here a simpler approach, based on
the following simplification: we consider that the variations of conductivity and
permittivity have a radial symmetry in the vicinity of the current sources.  This
simplification allows us to obtain simpler expressions of the extracellular
potential, still displaying frequency dependence, and apply this formalism using
standard simulation tools.

Consider Eq.~\ref{eq:FourPot} for the case of a spherically-symmetric system.
Then the potential obeys
\be
\label{eq:RadialFourPot}
\frac{d^{2}V_\omega}{dr^{2}} + \frac{2}{r} \frac{dV_\omega}{dr} +
\frac{1}{(\sigma+i\omega\epsilon)} \frac{d(\sigma+i\omega\epsilon)}{dr}
\frac{dV_\omega}{dr} = 0 ~ .
\ee
Integrating this equation gives the following relation between two points $r_1$
and $r_2$ in the extracellular space,
\be\label{eq:IntRadial}
r_1^{2} \ \frac{dV_\omega}{dr}(r_1) \
\left[ \sigma(r_1) + i \omega \epsilon(r_1) \right] =
r_2^{2} \ \frac{dV_\omega}{dr}(r_2) \
\left[ \sigma(r_2) + i \omega \epsilon(r_2) \right] ~ .
\ee
This can be verified by differentiating this equation with respect to $r$,
which then yields Eq.~\ref{eq:RadialFourPot}.  Integrating Eq.~\ref{eq:IntRadial}
once more yields
\be
\label{eq:IntIntRadial}
V_{\omega}(r_1) = V_{\omega}(r_2)
+ \frac{dV}{dr}(r_2) \int_{r_2}^{r_1} dr'
\frac{ r_2^{2} [\sigma(r_2) + i \omega~\epsilon(r_2)] }
{ r'^{2} [\sigma(r') + i \omega~\epsilon(r')] } ~ .
\ee
This can be seen most easily by differentiating Eq.~\ref{eq:IntIntRadial}
to yield Eq.~\ref{eq:IntRadial}.  In particular, if $r_2 = R$,
$-\frac{dV}{dr}(r_2)$ represents the electric field at the surface of
the sphere of radius $R$, which by Ohm's law is related to the current density
at $R$, we obtain
\be
\label{eq:VZI}
V_{\omega}(r_1) = V_{\omega}(R) - \frac{I_\omega}{4 \pi \sigma(R)}
\int_{R}^{r_1} dr' \ \frac{1}{r'^{2}} \
\frac{\sigma(R) + i \omega~\epsilon(R)}{\sigma(r') + i \omega~\epsilon(r')} ~ .
\ee
If we assume that the extracellular potential vanishes at large distances
($V_\omega(\infty)=0$), we have:
\be
\label{eq:Vinf}
V_{\omega}(\infty) = 0 = V_{\omega}(R) - \frac{I_\omega}{4 \pi \sigma(R)}
\int_{R}^{\infty} dr' \ \frac{1}{r'^{2}} \
\frac{\sigma(R) + i \omega~\epsilon(R)}{\sigma(r') + i \omega~\epsilon(r')} ~ .
\ee
which allows to eliminate $V_{\omega}(R)$ from Eq.~\ref{eq:VZI}, leading to:
\be
\label{eq:Vgen}
V_{\omega}(r_1) = \frac{I_\omega}{4 \pi \sigma(R)}
\int_{r_1}^{\infty} dr' \ \frac{1}{r'^{2}} \
\frac{\sigma(R) + i \omega~\epsilon(R)}{\sigma(r') + i \omega~\epsilon(r')} ~ .
\ee
This will be the main equation that forms the basis of our simplified model of
LFP.  Solving this equation for $\sigma$ and $\epsilon$ constant leads to the
expression found above (Eq.~\ref{eq:SpherSol}) for homogeneous extracellular
media.  In the numerical part (Section {\it Numerical simulations}), we will
solve this equation for different spatial profiles of $\sigma$ and $\epsilon$. 
To this end, it is useful to define the impedance:
\be
\label{eq:DefImped}
Z_\omega(r_1) = \frac{1}{4 \pi \sigma(R)}
\int_{r_1}^{\infty} dr' \ \frac{1}{r'^{2}} \
\frac{\sigma(R) + i \omega~\epsilon(R)}{\sigma(r') + i \omega~\epsilon(r')} ~ .
\ee
Then Eq.~\ref{eq:Vgen} becomes
\be
V_\omega(r_1) \ = \ Z_\omega(r_1) \ I_\omega ~ .
\ee
The impedance is therefore the ``filter'' applied to the $\omega$-frequency
component of the current source, to yield the corresponding frequency component
of the extracellular potential.  In the next section, we will examine the
frequency-filtering properties of different extracellular media by calculating
numerically the impedance for different cases of spatial inhomogeneities of
conductivity and permittivity.

\sect{NUMERICAL SIMULATIONS}
\label{sec:SimulResults}

In this section, we use the expressions of the extracellular potential obtained
above.  In particular we analyze the behavior of the extracellular potential
generated by a current source in a spherically-symmetric non-homogeneous medium
(Eq.~\ref{eq:Vgen}), and its associated impedance (Eq.~\ref{eq:DefImped}).
We investigated the frequency-filtering properties obtained for different
cases of increasing complexity of the radial profile of $\sigma$ and $\epsilon$.
The goal is to determine the conditions of spatial variations of conductivity
and permittivity for which the frequency-filtering properties are consistent
with physiological data.  We terminate by an application of this model to
calculating the LFP generated by a conductance-based spiking neuron model.

\subsect{Parameters}
\label{sec:PhysiolData}

Precise experimental data on the variations of permittivity $\epsilon$ and
conductivity $\sigma$ in the extracellular medium have not been measured so far. 
However, averaged values of these parameters are available from macroscopic
measurements.  A value for $\sigma$, averaged over large extracellular distances,
$\sigma_{av}$, was measured by Ranck (1963) and was between 0.28~$S/m$ and
0.43~$S/m$, for 5~Hz and 5~kHz, respectively.  The macroscopic frequency
dependence of conductivity seems therefore relatively weak.  However, the
situation is different microscopically.  As reviewed in Nunez (1981), the
conductivity of the CSF fluid is 1.56~$S/m$ while the typical conductivity of
membranes is $3.5 \times 10^{-9}$~$S/m$.  This value was obtained from the
resting (leak) membrane conductance of cortical neurons, typically around $4.5
\times 10^{-5}$~S/cm$^2$, multiplied by the thickness of the membrane (7-8~nm;
Peters et al., 1991).  At microscopic scales, there is therefore approximately 9
orders of magnitude variations of conductivity.

Permittivity variations are not so dramatic.  Fluids have higher permittivity,
for example it is about $7 \times 10^{-10}$~$F/m$ for sea water.  Membranes have
a permittivity of about $7.5 \times 10^{-11}$~$F/m$.  The latter value was
derived from the specific capacitance of membranes, $C$ = 1~$\mu$~F/cm$^2$
(Johnston and Wu, 1997), and assuming a membrane thickness of 7.5~nm (Peters et
al., 1991).  Because those variations are small compared to the variations of
conductivity, it is a good approximation to consider the permittivity as a
constant.  In the following, we will use the reference value of $\epsilon =
10^{-10}$~$F/m$.

%
%
%
%
%
%
%

In the following we will use normalized values for conductivity
$\sigma(r)/\sigma(R)$ and permittivity $\epsilon(r)/\sigma(R)$.  Because the
membrane is always surrounded by extracellular fluid, $\sigma(R) = 1.56$~$S/m$
and the normalized conductivity $\sigma(r)/\sigma(R)$ therefore varies between 1
and about $2 \times 10^{-9}$.  Similarly, the normalized (constant) value of
permittivity will be $\epsilon(r)/\sigma(R) = 6 \times 10^{-11}$~$s$.


\subsect{Frequency-filtering properties of spherically-symmetric media}
\label{sec:ImpedanceResults}

\begin{figure}[ht]
\insertc{\includegraphics[width=17cm]{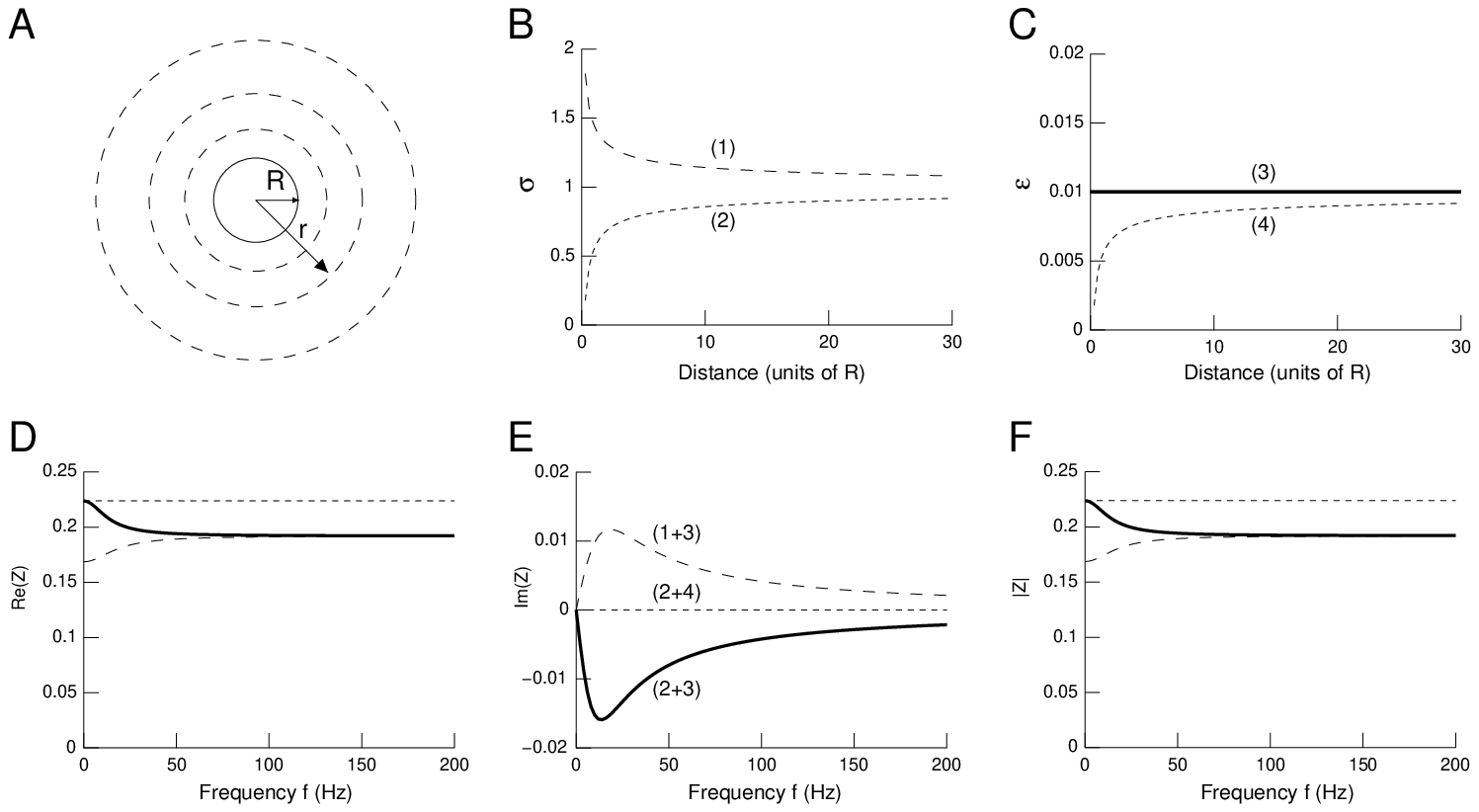}}
\caption{\label{simple} Radial variations of conductivity and permittivity
can induce frequency-filtering properties.}
A. Scheme of the current source in radial symmetry.  The current source is
assumed to be spherical (continuous line; radius $R$).  The conductivity and
permittivity vary in radial symmetry according to the distance $r$ from the
center of the source.
B. Conductivity $\sigma$ vs.\ radial distance $r$.  Two cases are shown: (1)
$\sigma(r)/\sigma(R) = 1 + \sqrt{r_{0}/r}$ and (2) $\sigma(r)/\sigma(R) = 1 -
\sqrt{r_{0}/r}$, where $r_{0}=0.2025~R$ ($R$=1 here).
C. Permittivity $\epsilon$ vs.\ radial distance $r$. The two curves shown are:
(3) $\epsilon(r)/\sigma(R)= 0.01$ and (4) $\epsilon(r)/\sigma(R) = 0.01 \ [1 -
\sqrt{r_{0}/r}]$.
D-F. Real part (D), imaginary part (E) and norm (F) of the impedance
$Z_{\omega}(r=5R)$ vs.\ frequency $f$.  Combining the profiles (1) and (3) in
B-C leads to a high-pass filter (dashed line), whereas (2+3) gives low-pass
characteristics (solid line).  The combination (2+4) is such that
$\sigma(r)/\epsilon(r) = const.$, in which case there is no frequency
dependence (dotted line).
\end{figure}

We calculated numerically the impedance (Eq.~\ref{eq:DefImped}) for
different cases of spatial variations of conductivity and permittivity.
In all cases we assumed a current source with spherical geometry,
characterized by radius $R$, and that $\sigma$ and $\epsilon$ vary
according to a radial (spherical) symmetry around this current source
(see scheme in Fig.~\ref{simple}A).  For each case, we represented the
normalized impedance
\be
\tilde{Z}_\omega(r) = Z_\omega(r) / Z_\omega(R) ~ ,
\ee
which allows better comparison between the different cases.  Because the value of
impedance does not depend on the absolute value of permittivity and conductivity
(dividing $\sigma$ and $\epsilon$ by a constant factor does not change
$Z_\omega(r)$ in Eq.~\ref{eq:DefImped}), we used the normalized conductivity
$\sigma(r)/\sigma(R)$ and the normalized permittivity $\epsilon(r)/\sigma(R)$
defined above.  It is also convenient to represent all distances in units of $R$,
although we also considered absolute values of distances (see below).

We first investigated a simple case of smooth variations of those parameters, to
illustrate the different types of frequency filtering that can be obtained in
this model.  The profiles of conductivity and permittivity are shown in
Figs.~\ref{simple}B and C.  These curves tend to the same asymptotic value for
large distances.  The corresponding impedance is shown as a function of frequency
$f$ in Figs.~\ref{simple}D-F (see Appendix~3 for details of the method).  When
the ratio $\sigma/\epsilon$ is kept constant (Fig.~\ref{simple}D-F, dotted line),
there is no frequency dependence as analyzed above in Section {\it Special
cases}.  In the case of a decreasing conductivity with distance combined with
constant permittivity, one has a high-pass filter (Fig.~\ref{simple}D-F, dashed
line).  By contrast, a low-pass filter is observed if an increasing conductivity
with distance is combined with a constant permittivity (Fig.~\ref{simple}D-F,
solid line).  Thus, there is a clear frequency-dependent behavior when $\sigma$
and/or $\epsilon$ vary as function of distance $r$, if the ratio
$\sigma/\epsilon$ is not constant.  This also shows that low- and high-pass
filters are both possible, depending on the exact form of the function
$\sigma(r)$ and $\epsilon(r)$.  The impedance can also have a non-zero imaginary
part, which means that beyond resistivity, the medium has is also capacitive
properties.  In this case, there will be a phase difference between the potential
and the current.

\bigskip

\begin{figure}[ht]
\insertc{\includegraphics[width=17cm]{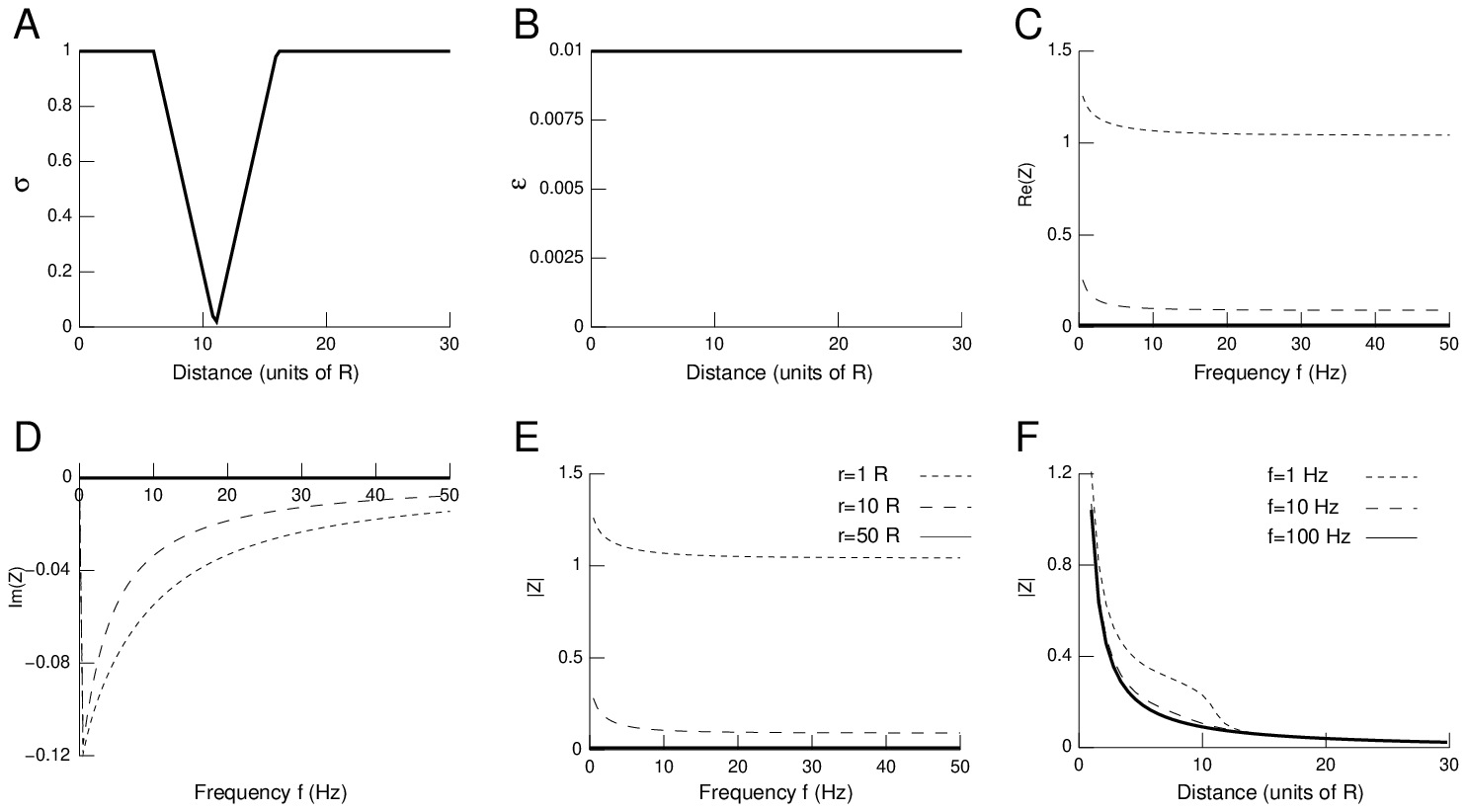}}
\caption{\label{drop} Frequency-filtering properties obtained by a localized
drop in conductivity.}
A. Profile of conductivity vs.\ distance.  The conductivity was described by
$\sigma(r)/\sigma(R) = 1 - 0.2 \ (r-6R)/R$ for $6R < r < 11R$,
$\sigma(r)/\sigma(R) = -1 + 0.2 \ (r-6R)/R$ for $11R < r < 16R$,
and $\sigma(r)/\sigma(R) = 1$ otherwise.
B. Profile of permittivity.  $\epsilon(r)/\sigma(R)$ was constant and equal to
0.01.
C-E.  Real part (C), imaginary part (D) and norm (E) of the impedance as a
function of frequency $f$.  $Z_{\omega}(r)$ is shown for different distances $r$
away from the source.
F. Attenuation of the impedance norm $|Z_\omega(r)|$ with distance. The
different curves correspond to three different frequencies.
\end{figure}

We next considered a case characterized by a localized drop of conductivity
(Fig.~\ref{drop}A) while permittivity was kept constant (Fig.~\ref{drop}B).  The
resulting impedance measured at different distances from the source is shown in
Fig.~\ref{drop}C-E as a function of frequency $f$.  In this case, for distances
around the conductivity drop, there is a moderate frequency dependence with
low-pass characteristics (Fig.~\ref{drop}C-E, dotted and dashed lines).  However,
for larger distances, the imaginary part is zero and there is no frequency
dependence (Fig.~\ref{drop}C-E, solid lines).  This is explained by the fact that
for large distances $\sigma(r)=\sigma(R)$ and $\epsilon(r)=\epsilon(R)$.  This
behavior can also be seen in the attenuation of the different frequency
components illustrated in Fig.~\ref{drop}F.  There is a different attenuation
only for distances around the region where conductivity varies.

\bigskip

\begin{figure}[ht]
\insertc{\includegraphics[width=17cm]{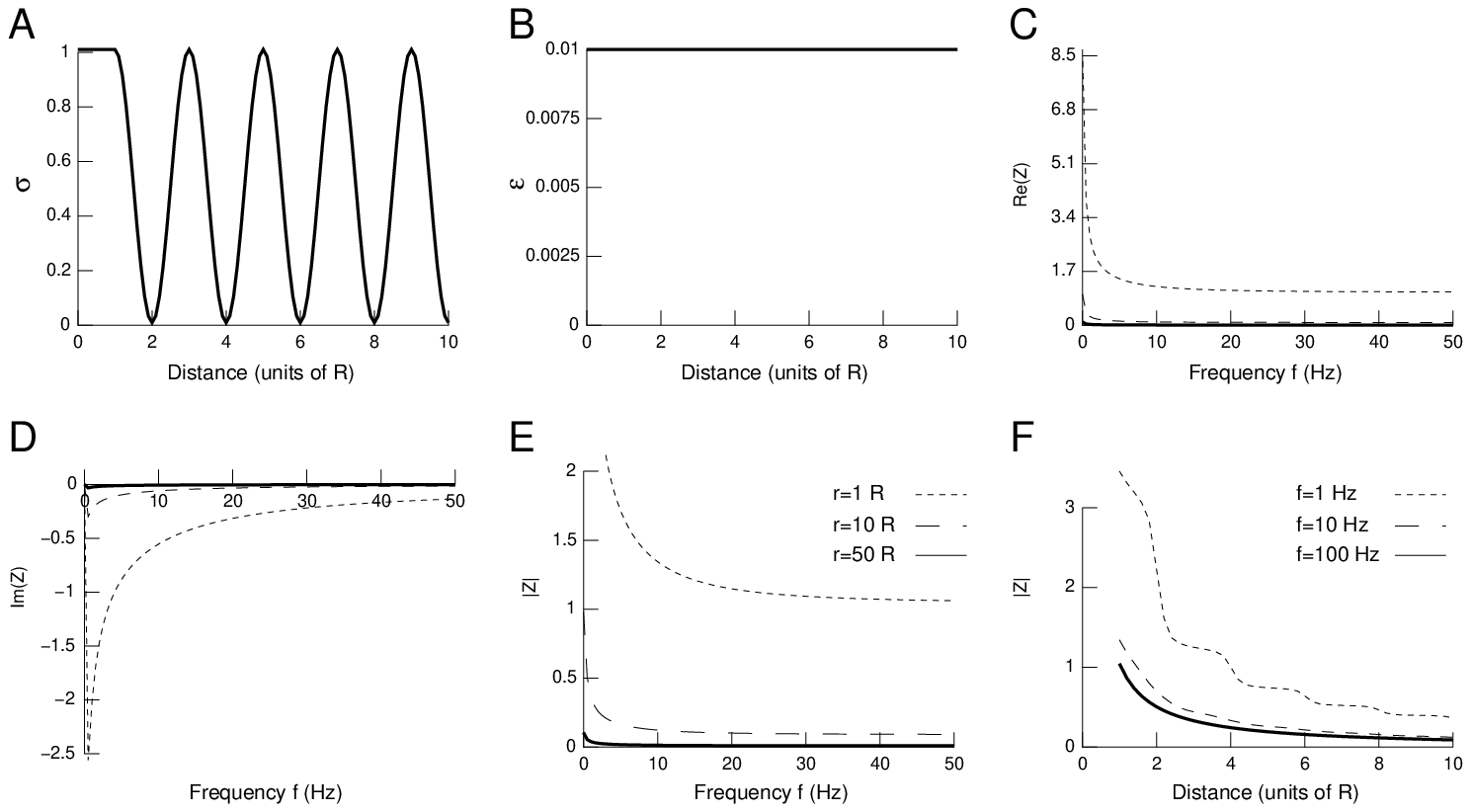}}
\caption{\label{osc} Frequency-filtering properties obtained from a periodically
varying conductivity.}
A. Oscillatory profile of conductivity vs.\ distance ($\sigma(r)/\sigma(R) =
0.501 + 0.5 * \cos[2 \pi (r-R) / 2R]$.
B. Profile of permittivity ($\epsilon(r)/\sigma(R)$ = 0.01).
C-E. Real part (C), imaginary part (D) and norm (E) of the impedance
$Z_{\omega}(r)$ vs.\ frequency $f$.  The different curves are taken at different
distances $r$ outside of the current source.
F. Attenuation of the impedance norm $|Z_\omega(r)|$ with distance.  The
different curves indicate the attenuation obtained at different frequencies.
\end{figure}

Because the extracellular space is composed of alternating fluids and membranes
(Peters et al., 1991), which have high and low conductivity, respectively, we
have next considered the situation where conductivity fluctuates periodically
with distance (Fig.~\ref{osc}).  Considering a cosine function of conductivity
(Fig.~\ref{osc}A) with constant permittivity (Fig.~\ref{osc}B) leads to a rather
strong frequency-dependent attenuation (Fig.~\ref{osc}C-E) with low-pass
characteristics.  There was a strong attenuation with distance for all
frequencies (Fig.~\ref{osc}F).  Very similar results were obtained with other
periodic functions (for example by replacing cos by sin in the function used in
Fig.~\ref{osc}A), different oscillation periods, or even for damped oscillations
of conductivity (not shown).

\bigskip

\begin{figure}[ht] 
\insertc{\includegraphics[width=17cm]{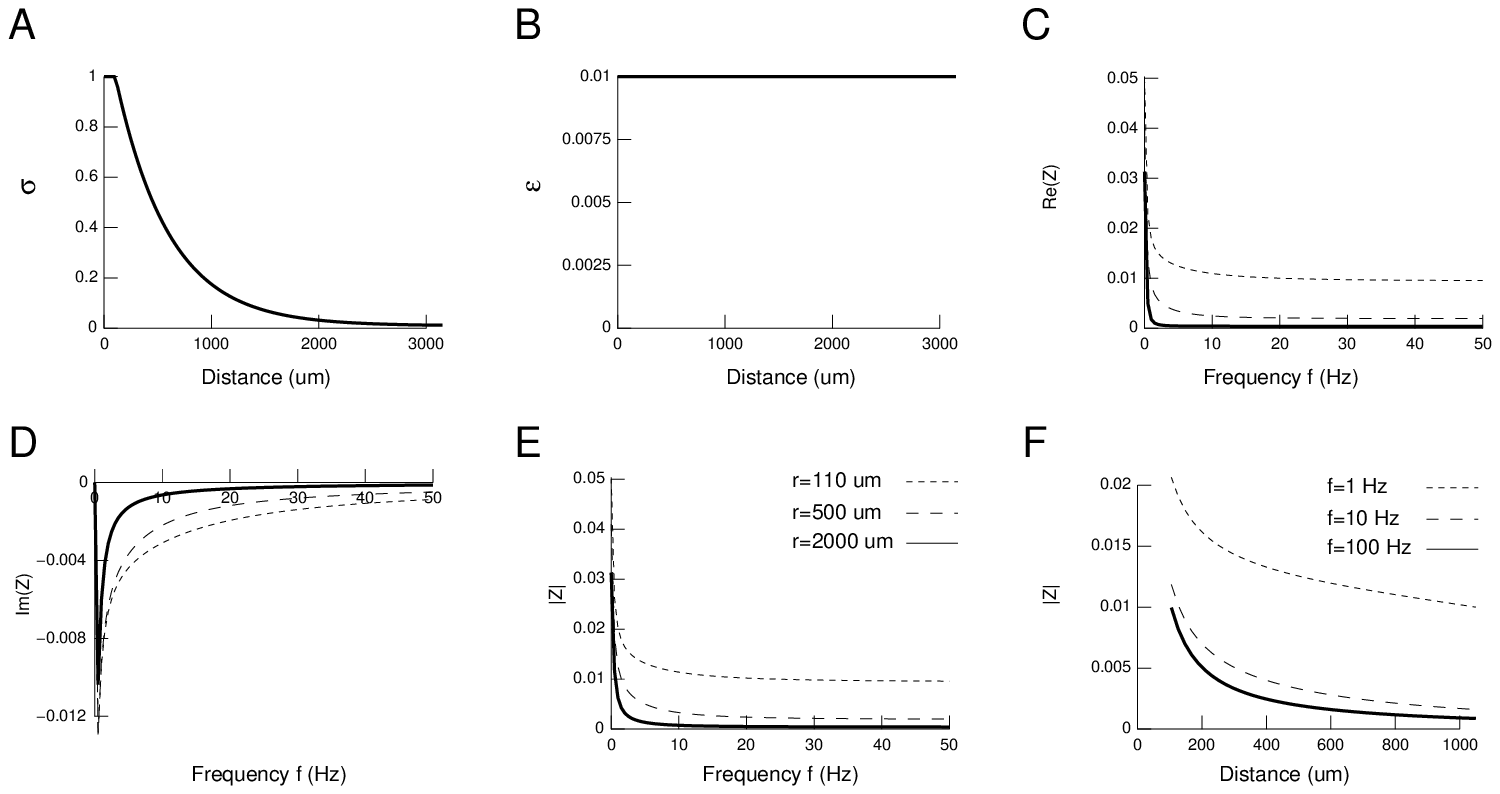}}
\caption{\label{exp} Frequency-filtering properties obtained with exponential
decrease of conductivity.}
A. Profile of conductivity.  $\sigma(r)/\sigma(R)$ decays exponentially according
to $\sigma(r)/\sigma(R)$ = $\sigma_0 + (1-\sigma_0) \ \exp[-(r-R)/\lambda]$,
with a space constant $\lambda=500~\mu$m.
B. Profile of permittivity.  $\epsilon(r)/\sigma(R)$ was constant (0.01).
C-E. Real part (C), imaginary part (D) and norm (E) of the impedance
$Z_{\omega}(r)$ vs.\ frequency $f$.  The different curves show the impedance
calculated at different distances $r$.
F. Attenuation of the impedance norm $|Z_\omega(r)|$ with distance.  The
different curves indicate the attenuation obtained at different frequencies.
\end{figure}

It could be argued that although fluids and membranes alternate in extracellular
space, there is an efficient diffusion of ions only in the extracellular fluid
around the membrane.  For larger distances, diffusion becomes increasingly
difficult because of the increased probability of meeting obstacles.  In this
case, conductivity would be highest around the source and progressively decrease
to an ``average'' conductivity level for larger distances.  This situation is
illustrated in Fig.~\ref{exp}.  We have considered that the conductivity is
highest at the source, then decreases exponentially with distance with a space
constant $\lambda$ (Fig.~\ref{exp}A; note that in this case, real distances were
used).  Permittivity was constant (Fig.~\ref{exp}B).  The resulting impedance
displayed pronounced frequency-filtering properties with low-pass characteristics
(Fig.~\ref{exp}C-E).  In particular, the attenuation with distance revealed
strong differences between low and high frequencies of the spectrum
(Fig.~\ref{exp}F).  Similar results can be obtained with other decreasing
functions of connectivity (not shown).

\bigskip

\begin{figure}[ht] 
\insertc{\includegraphics[width=17cm]{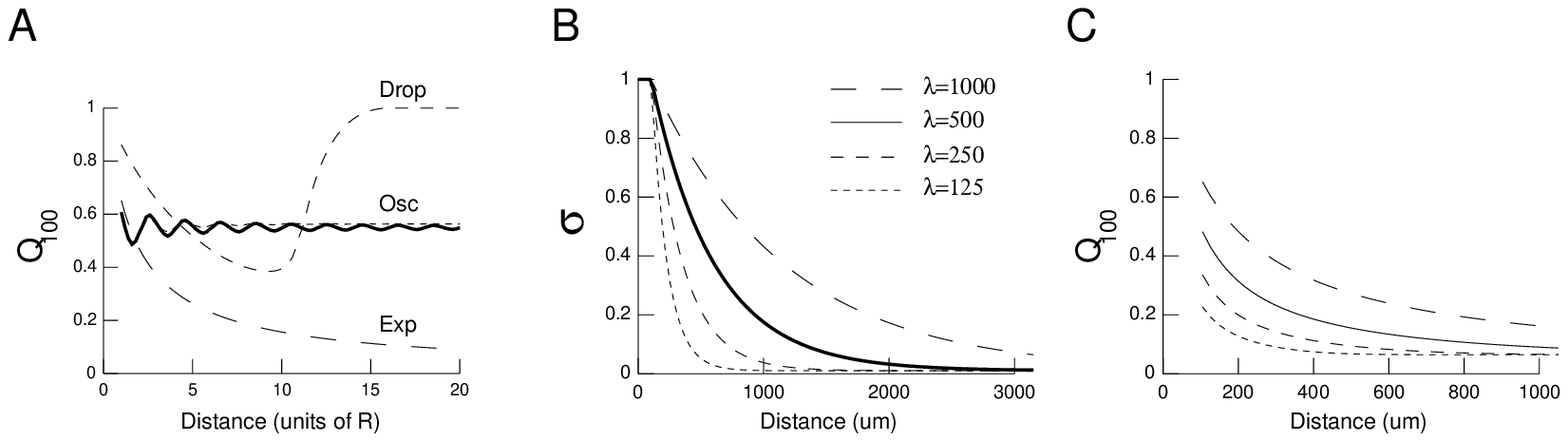}}
\caption{\label{ratio} Distance dependence of frequency-filtering properties}
A. Ratio of impedance at fast and slow frequencies ($Q_{100}$) represented as a
function of distance $r$ (units of $R$).  The $Q_{100}$ ratios are represented
for different profiles of conductivity. {\it Drop}: localized drop of
conductivity (short dash; same parameters as in Fig.~\ref{drop}).  {\it Osc}:
oscillatory profile of conductivity (solid line; same parameters as in
Fig.~\ref{osc}; the dotted line indicates a damped cosine oscillation).  {\it
Exp}: exponential decrease of conductivity (long dash; same parameters as in
Fig.~\ref{exp} except $R=1$, $\lambda$=10R).
B. Profiles of conductivity with exponential decay (same parameters as in
Fig.~\ref{exp}; space constants $\lambda$ indicated in $\mu$m).
C. $Q_{100}$ ratios obtained for the conductivity profiles shown in B.
\end{figure}

The above examples show that there can be a strong frequency-filtering behavior,
with low-pass characteristics as observed in experiments.  However, although
these examples show a more effective filtering for high frequencies, it still
remains to be shown that the high frequencies attenuate more steeply with
distance compared to low frequencies. To this end, we define the quantity:
\be
Q_{100} \ = \ Z_{100}(r) / Z_{1}(r) ~ ,
\ee
where $Z_1$ and $Z_{100}$ are the impedances computed at 1~Hz and 100~Hz,
respectively. This ratio quantifies the differential filtering of fast and slow
frequencies as a function of distance $r$.  Fig.~\ref{ratio}A displays the
$Q_{100}$ values obtained for some of the examples considered above.  In the case
of a localized drop of conductivity (Fig.~\ref{ratio}A, {\it Drop}), there was an
effect of distance for $r<16R$, then the $Q_{100}$ remained equal to unity for
further distances.  This behavior is in agreement with the impedance shown in
Fig.~\ref{drop}, in which case there was no frequency filtering for $r>16R$.  For
oscillatory conductivities (Fig.~\ref{ratio}A, {\it Osc}), the $Q_{100}$ was
always $<1$, consistent with the low-pass frequency-filtering behavior observed
in Fig.~\ref{osc}. However, the $Q_{100}$ oscillated around a value of 0.6 and
did not further decrease with distance.  Thus, in this case, although there was a
clear low-pass filtering behavior, all frequencies still contribute by the same
relative amount to the extracellular potential, regardless of distance. On the
other hand, with exponential decay of conductivity, the $Q_{100}$ monotonically
decreased with distance (Fig.~\ref{ratio}A, {\it Exp}). Thus, this case shows
both low-pass filtering behavior (Fig.~\ref{exp}) and a stronger attenuation
of high frequencies compared to low frequencies (Fig.~\ref{ratio}A, {\it
Exp}), which is in qualitative agreement with experiments. Analyzing
exponentially-decaying conductivities of different space constants
(Fig.~\ref{ratio}B) revealed that the various patterns of distance dependence
approximately followed the pattern of conductivity (Fig.~\ref{ratio}C). This
type of conductivity profile is relatively simple and plausible, and will be the
one considered in the biophysical model investigated below.

\subsect{Biophysical model of the frequency-filtering properties of
local field potentials}
\label{sec:BiophysModel}

\begin{figure}[ht]
\insertc{\includegraphics[width=15cm]{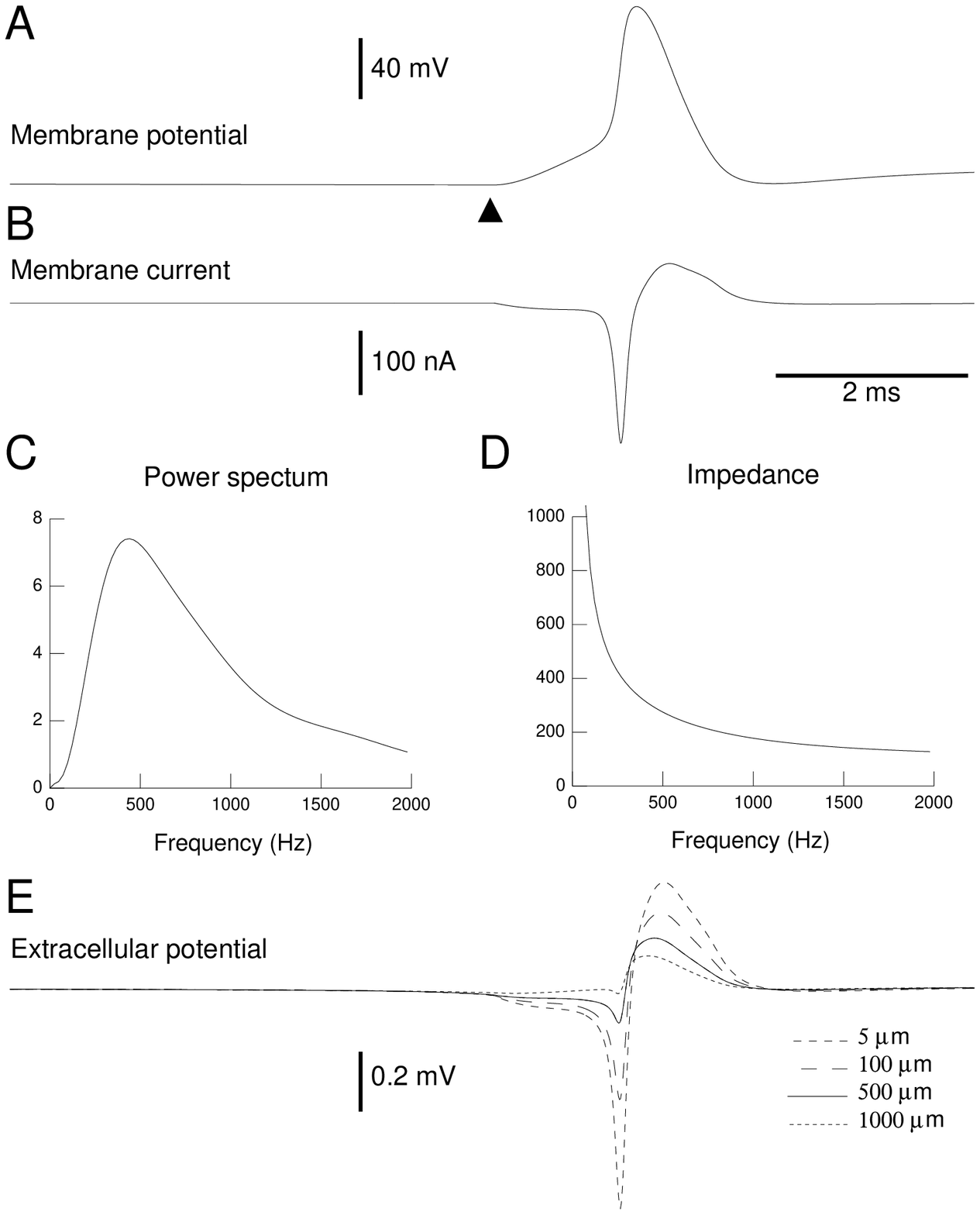}}
\caption{\label{nrn} Frequency-filtered extracellular field potentials in a
conductance-based model.}
A. Membrane potential of a single-compartment model containing voltage-dependent
Na$^+$ and K$^+$ conductances and a glutamatergic synaptic conductance.  The
glutamatergic synapse was stimulated at $t=5~ms$ (arrow) and evoked an action 
potential. 
B. Total membrane current generated by this model.  Negative currents correspond
to Na$^+$ and glutamatergic conductances (inward currents), while positive
currents correspond to K$^+$ conductances (outward currents).
C. Power spectrum of the total current shown in B.  
D. Impedance at 500~$\mu$m from the current source assuming a radial profile of 
conductivity and permittivity (as in Fig.~\ref{exp}).  
E. Extracellular potential calculated at various distances from the source (5,
100, 500 and 1000~$\mu$m).  The frequency filtering properties can be seen by
comparing the negative and positive deflections of the extracellular potential. 
The fast negative deflection almost disappeared at 1000~$\mu$m whereas the slow 
positive deflection was still present.
\end{figure}

We have applied the above formalism to model the frequency dependence of the
extracellular field potentials stemming from a conductance-based spiking neuron
model.  The details about the model are given in the {\it Material and methods}
section, while the details of the calculation of the extracellular LFP (general
for any current source) is given in Appendix~3. The profile of conductivity and
permittivity used is that of Fig.~\ref{exp}.  We calculated the total membrane
current generated by a single-compartment model of an adapting cortical neuron,
containing voltage-dependent Na$^+$ and K$^+$ conductances for generating action
potentials and a slow voltage-dependent K$^+$ conductance responsible for
spike-frequency adaptation.  The model also contained a fast glutamatergic
excitatory synaptic conductance, which was adjusted to evoke a post-synaptic
potential just above threshold, in order to evoke a single action potential
(Fig.~\ref{nrn}A).  The total membrane current (Fig.~\ref{nrn}B) was calculated
and stored in order to calculate its Fourier transform (power spectral density
shown in Fig.~\ref{nrn}C).  The impedance of the extracellular medium
(Fig.~\ref{nrn}D) was calculated using absolute values of the parameters
(Eq.~\ref{eq:DefImped}).  

This model was used to calculate the field potentials at different radial
distances assuming the neuron was a spherical source (radius of 105~$\mu$m).  The
extracellular potential is indicated for 5, 100, 500 and 1000~$\mu$m away from
the source (see Fig.~\ref{nrn}E) and strong frequency filtering properties are
apparent: the fast negative deflection of extracellular voltage showed a steep
attenuation and almost disappeared at 1000~$\mu$m (although it had the highest
amplitude at 5~$\mu$m).  In contrast, the slow positive deflection of the
extracellular potential showed less attenuation with distance and became dominant
at large distances (500 and 1000~$\mu$m in the example of Fig.~\ref{nrn}E). 

Thus, this simple example illustrates that the approach provided here can lead to
a relatively simple model to calculate local field potentials with frequency
filtering properties.  The exact profiles of filtering and attenuation depend on
the exact shape of the gradients of conductivity/permittivity as well as on the
spherical symmetry inherent to this model.


\sect{DISCUSSION}
\label{sec:Discuss}

In this paper, we have provided a model of extracellular field potentials
in non-homogeneous media.  We discuss here the validity of this model, how
it relates to previous studies, and what perspectives are provided.

The theoretical analysis outlined in Section {\it General theory} shows that
inhomogeneities of extracellular space (with respect to conductivity and/or
permittivity) is a possible cause for frequency filtering.  In general,
non-homogeneous extracellular media will differently affect the attenuation of
the various frequency components of the current sources, and can lead to
high-pass or low-pass filters depending on the gradients of conductivity and
permittivity. The composition of extracellular space is made from the alternance
of fluids and membranes (Peters et al., 1991).  Because these media have very
different conductivity and permittivity, one may expect that the extracellular
space is necessarily highly non-homogeneous.  Therefore, the structural
composition of extracellular space is very likely to be a main determinant of the
frequency-filtering properties of LFPs.  In addition, the conductivity of the
extracellular fluid directly beneath the membrane depends on the ionic
concentrations present.  It turns out that the extracellular ionic concentrations
may vary in time, in an activity-dependent manner (reviewed in Amzica, 2002).
Therefore, it is also likely that there is an activity dependent contribution to
the filtering properties of the extracellular medium.  Here, we did not consider
such time-dependent variations of conductivity, but this type of contribution is
certainly worth to be considered by future theoretical work.

To correctly simulate the frequency-filtering behavior due to extracellular
inhomogeneity, the extracellular potential should be calculated by a model
incorporating details about the three-dimensional composition of the
extracellular medium.  Such type of simulations should use methods such as
finite-element analysis.  However, the complexity of this type of analysis,
and of the data it requires, makes such simulations inaccessible to standard
models.  In addition, this requires orders of magnitude differences in
computational power needs.  For these reasons, we have considered the option
of generating a simplified model under some approximation.  We assumed that
the geometry of extracellular inhomogeneities is spheric around the current
source.  In this condition, one can obtain relatively simple expressions of
the extracellular potential such as Eq.~\ref{eq:Vgen}.  Not only this expression
is amenable to theoretical analysis, but it is also sufficiently simple to be
applied to current neuron models which do not have an explicit representation
of extracellular space.

The drawback of this method is that it considers an un-realistic (radial)
distribution of inhomogeneities in extracellular space, which will necessarily
affect the frequency-filtering properties produced by the model.  However, it
should be possible to calculate the ``average'' radial variations of conductivity
and permittivity by averaging the profiles of $\sigma$ and $\epsilon$ in all
directions emanating from neuronal membranes using three-dimensional
reconstructions of the neuropil.  Another direction would be to measure
experimentally the profiles and distributions of $\sigma$ and $\epsilon$, but
such data are not currently available.  Here, we have used ``heuristic'' profiles
of conductivity and permittivity, which gives frequency-filtering properties in
qualitative agreement with experiments.  It should be easy to make this model
more realistic by incorporating different radial functions of $\sigma$ and
$\epsilon$ when there will be better constraints by measurements.

The present results lead to several interesting perspectives for future work or
extensions.  First, as mentioned above, the simplified model could be enhanced by
comparison with a more realistic model, for example based on three-dimensional
reconstructions of extracellular space.  The simplified model could be adjusted
so that it fits as closely as possible the behavior of the more realistic model,
yielding more optimal expressions of the radial profiles of conductivity and
permittivity.  A second possible direction is the ``reverse'' problem of
estimating neuronal activity based on LFP measurements.  By using data on the
spatial and temporal variations of multisite LFPs and multi-unit activity, it
should be possible to estimate what are the respective contributions of the
natural frequency-filtering properties of extracellular space and the spatial
coherence of neuronal sources in the different frequency components.  For
example, it was shown that a consistent relation between LFP and cell firing
extends to large cortical distances ($>$7~mm) for slow-waves but not for fast
oscillations in the gamma (20-60~Hz) frequency range (Destexhe et al., 1999).  A
model of LFP is needed to evaluate whether this effect is really due to
differences in the coherence of neuronal firing (as the single-unit data
indicates), or if a large part could be explained by the low-pass filtering
properties of the extracellular medium.  A combination of experimental recordings
and computational models will be needed to understand how neuronal activity
translates into extracellular field potentials and vice-versa.


\sect{Acknowledgments}

H.K. has been supported by the National Science and Engineering Research Council
(NSERC, Canada). A.D. has been supported by the Medical Research Council (MRC,
Canada) and the Centre National de la Recherche Scientifique (CNRS, France).
Supplementary information is available at http://cns.iaf.cnrs-gif.fr

\insertp{\pagebreak}

\sect{REFERENCES}


\insertp{\pagebreak}
\appendix
\sect{APPENDIX}

\subsect{Appendix 1: Extracellular potential in non-homogeneous media with 
spherical symmetry}
\label{appendix1}

This Appendix refers to Section {\it Stationary currents in spherically-symmetric
non-homogeneous medium}: \\
Recall $\sigma=\sigma(r)$, $\epsilon=$const.
Then Gauss' law becomes
\be
\vec{\nabla} \cdot \vec{E} = \frac{\rho}{\epsilon} ~ .
\ee
The law of charge conservation becomes
\bea -\frac{\partial \rho}{\partial t} &=& \vec{\nabla} \cdot
(\sigma \vec{E}) = \sigma \vec{\nabla} \cdot \vec{E} + \vec{E}
\cdot (\vec{\nabla} \sigma)
\nonumber \\
&=& \frac{\sigma}{\epsilon} \rho + \vec{E} \cdot (\vec{\nabla} \sigma) ~ .
\eea
Using Ohm's law and assuming $\sigma \neq 0$, this becomes
\be \label{eq:ChargeEvol} -\frac{\partial \rho}{\partial t} =
\frac{\sigma}{\epsilon} \rho + \vec{j} \cdot (\frac{1}{\sigma}
\vec{\nabla} \sigma) =  \frac{\sigma}{\epsilon} \rho +  \vec{j}
\cdot \vec{\nabla} \log \sigma ~ . \ee

\bigskip

We make the assumption that the current density $\vec{j}$ is stationary, i.e.
it does not explicitly depend on time. Also, by assumption $\sigma$ and
$\epsilon$ are time independent.
Hence denoting $a= \frac{\sigma}{\epsilon}$ and
$b=\vec{j} \cdot \vec{\nabla} \log \sigma$,
Eq.~\ref{eq:ChargeEvol} takes the form
\be \frac{\partial \rho}{\partial t} = -a \rho - b  =0. \ee
Consequently, the induced charge density (at steady-state) is given by
\be \rho = - \frac{b}{a} ~ , \ee
\be
\rho = - \epsilon \vec{E} \cdot \vec{\nabla} \log \sigma ~ .
\ee
It shows that the net charge density is different from zero.

\bigskip

\noindent Electric field for spherical current source. \\
At the radius $r=R$, the electric field is given by the current
\be
\vec{j}|_{r=R} = \sigma(R) ~ \vec{E}|_{r=R} ~ .
\ee
The total current passing through a sphere (surface $S$, radius $r$) is given
by
\be
I = \int_{S} d\vec{S} \cdot \vec{j} = S(r) ~ j(r) ~ .
\ee
Thus the current density as function of radius $r$ behaves as
\be
\vec{j}(r) = j(r) ~ \vec{e}_{r} = \frac{I}{S(r)} ~ \vec{e}_{r} =
\frac{I}{4 \pi r^{2}} ~ \vec{e}_{r} ~ .
\ee
Ohm's law implies for the electric field
\be
\label{eq:ElecField}
\vec{E}(r) = \frac{1}{\sigma(r)} ~ \vec{j}(r) =
\frac{I}{4 \pi r^{2} \sigma(r)} ~ \vec{e}_{r} ~ .
\ee

\bigskip

\noindent Electric potential. \\
Because the electric field is radially symmetric, so is also the potential,
which obeys
\be
E(r) = - \frac{\partial}{\partial r} V(r) ~ .
\ee
Its solution is obtained from Eq.~\ref{eq:ElecField},
\be \label{eq:SolutionPot} V(r) = -\int_{\infty}^{r} dr' ~ E(r') =
\int_{r}^{\infty} dr' ~ \frac{I}{4 \pi r'^{2} \sigma(r')} ~ . \ee

\bigskip

\noindent An independent check of this solution can be obtained by considering
the law of charge conservation.
\be \vec{\nabla} \cdot (\sigma \vec{E}) = -\frac{\partial
\rho}{\partial t} ~ . \ee
Because of the time-independent charge density, this becomes
\be
\vec{\nabla} \cdot (\sigma \vec{\nabla} V ) =
(\vec{\nabla} \sigma) \cdot \vec{\nabla} V
+ \sigma \Delta V = 0 ~ .
\ee
Radial symmetry implies the following differential equation in the variable $r$,
\be
( \partial_{r} \sigma(r) )(\partial_{r} V(r) ) + \sigma(r)
\left[ \frac{2}{r} \partial_{r} V + \partial_{r}^{2} V \right] = 0 ~ .
\ee
A straight forward calculation shows that the potential, given by
Eq.~\ref{eq:SolutionPot} is the solution of this differential equation for $V
\to 0$ when $r \to \infty$.

\bigskip

\subsect{Appendix 2: Fourier component of the extracellular field potential}
\label{appendix2}

This Appendix refers to Section {\it Time-varying currents in non-homogeneous
medium}: \\
Starting from Gauss' law
\be
\vec{\nabla} \cdot (\epsilon \vec{E}) = \rho ~ ,
\ee
the inhomogeneity of $\epsilon$ implies
\be
\vec{E} \cdot (\vec{\nabla} \epsilon) + \epsilon \vec{\nabla} \cdot \vec{E} =
\rho ~ .
\ee
Then the electric potential obeys
\be
- (\vec{\nabla} V) \cdot (\vec{\nabla} \epsilon) - \epsilon \Delta V = \rho ~ .
\ee
We recall that the potential $V$ and the charge density $\rho$ are
time-dependent, while the permittivity $\epsilon$ is not.
We define the Fourier transform of time-dependent function $f(t)$ via
\be f_{\omega} = \int_{-\infty}^{\infty} dt ~ e^{i \omega t} ~
f(t) ~ . \ee
Now we perform a Fourier transform with respect to time of the potential $V$
and the charge density $\rho$ to obtain an equation for the Fourier components
at frequency $\omega$,
\be
\label{eq:CompPotEps}
(\vec{\nabla} V_{\omega}) \cdot (\vec{\nabla} \epsilon) + \epsilon
\Delta V_{\omega} = - \rho_{\omega} ~ .
\ee

\bigskip

Similarly, starting from the differential form of the law of charge
conservation,
\be \vec{\nabla} \cdot (\sigma \vec{E}) = -\frac{\partial
\rho}{\partial t} ~ , \ee
the inhomogeneity of $\sigma$ implies
\be \vec{E} \cdot (\vec{\nabla} \sigma) + \sigma \vec{\nabla}
\cdot \vec{E} = -\frac{\partial \rho}{\partial t} ~ . \ee
Then the potential obeys
\be (\vec{\nabla} V) \cdot (\vec{\nabla} \sigma) + \sigma \Delta V
= \frac{\partial \rho}{\partial t} ~ . \ee
From the Fourier transform of $\rho$ follows that a component at frequency
$\omega$ satisfies
\be \frac{\partial}{\partial t} \rho_{\omega} = i \omega
\rho_{\omega} ~ . \ee
The last two equations imply that the Fourier component at frequency $\omega$
of the potential obeys
\be
\label{eq:CompPotSigm}
(\vec{\nabla} V_{\omega}) \cdot (\vec{\nabla} \sigma) + \sigma \Delta
V_{\omega} = - i \omega \rho_{\omega} ~ .
\ee
Combining Eqs.~\ref{eq:CompPotEps} and \ref{eq:CompPotSigm} yields
\be
\label{eq:CompPotA}
(\vec{\nabla} V_{\omega}) \cdot (\vec{\nabla} (\sigma + i \omega \epsilon)) +
 (\sigma + i \omega \epsilon) \Delta V_{\omega} = 0 ~ .
\ee

\subsect{Appendix 3: Method to calculate the extracellular field potential 
from point current sources}
\label{appendix3}

\begin{enumerate}

\item Compute the Fourier component $\omega$ of the impedance
\be
\label{eq:Imped3}
Z_\omega(r) = \frac{1}{4 \pi \sigma(R)}
  \int_{r}^{\infty} dr' \ \frac{1} {r'^2} \
                     \frac{ [\sigma(R) + i \omega~\epsilon(R)] }
                          { [\sigma(r') + i \omega~\epsilon(r')] } ~ ,
\ee
where $\omega = 2 \pi f$.  This expression incorporates the values of the
conductivity $\sigma(r)$ and permittivity $\epsilon(r)$ as a function
of the distance $r$.  It is also assumed that $V(\infty)=0$.

This quantity is computed for each frequency component $\omega$ of the
spectrum, and for each extracellular distance $r$ considered.  It can be
precalculated and stored in a matrix ($Z[f][r]$).

\item For each current source, compute the (complex) Fourier transform of the
total membrane current, which we call here $I_\omega$.

\item For each current source, compute the Fourier component $\omega$ of
the extracellular potential:
\be
\label{eq:FourV}
V_\omega(r) \ = \ Z_\omega(r) \ I_\omega  ~ .
\ee


\item For each current source, compute the extracellular potential by
applying the (complex) inverse Fourier transform to Eq.~\ref{eq:FourV}.

\item Finally, combine the contributions from all current sources to yield
the extracellular potential at a given position $\vec{x}$ in the
extracellular space.

\end{enumerate}



\clearpage
\sect{FIGURES}


\insertc{\clearpage}


\insertc{\end{document}}	

\clearpage
\includegraphics[width=17cm]{simple.ps}
\center{Figure~\ref{simple}}

\clearpage
\includegraphics[width=17cm]{drop.ps}
\center{Figure~\ref{drop}}

\clearpage
\includegraphics[width=17cm]{cosinus.ps}
\center{Figure~\ref{osc}}

\clearpage
\includegraphics[width=17cm]{exp2.ps}
\center{Figure~\ref{exp}}

\clearpage
\includegraphics[width=17cm]{ratios.ps}
\center{Figure~\ref{ratio}}

\clearpage
\includegraphics[width=15cm]{biophys.ps}
\center{Figure~\ref{nrn}}


\begin{enumerate}

\bibitem{Amzica2002} Amzica, F. 2002. In vivo electrophysiological evidences for
cortical neuron-glia interactions during slow ($<$1~Hz) and paroxysmal sleep
oscillations.  J. Physiol. Paris 96: 209-219.

\bibitem{Bremer38b} Bremer, F. 1938. L'activit\'e \'electrique de l'\'ecorce
c\'er\'ebrale.  Actualit\'es Scientifiques et Industrielles 658: 3-46.

\bibitem{Bremer49} Bremer, F. 1949)  Consid\'erations sur l'origine et la nature
des ``ondes'' c\'er\'ebrales.  Electroencephalogr. Clin.  Neurophysiol. 1:
177-193.

\bibitem{Creutzfeldt66a} Creutzfeldt, O., S. Watanabe, and H.D. Lux. 1966a. 
Relation between EEG phenomena and potentials of single cortical cells.  I. 
Evoked responses after thalamic and epicortical stimulation.  Electroencephalogr.
Clin.  Neurophysiol. 20: 1-18.

\bibitem{Creutzfeldt66b} Creutzfeldt, O., S. Watanabe, and H.D. Lux. 1966b. 
Relation between EEG phenomena and potentials of single cortical cells.  II. 
Spontaneous and convulsoid activity. Electroencephalogr. Clin. Neurophysiol. 20:
19-37.

\bibitem{Destexhe98} Destexhe, A. 1998. Spike-and-wave oscillations based on the
properties of GABA$_B$ receptors. J. Neurosci. 18: 9099-9111.

\bibitem{Destexhe99} Destexhe, A., D. Contreras, M.Steriade. 1999. 
Spatiotemporal analysis of local field potentials and unit discharges in cat
cerebral cortex during natural wake and sleep states.  J. Neurosci. 19: 4595-4608.

\bibitem{Noisy98} Destexhe, A. and D. Par\'e. 1999.  Impact of network activity
on the integrative properties of neocortical pyramidal neurons in vivo.  J. 
Neurophysiol. 81: 1531-1547.


\bibitem{Eccles51} Eccles, JC. 1951.  Interpretation of action potentials evoked
in the cerebral cortex.  J. Neurophysiol. 3: 449-464.

\bibitem{Henze2000} Henze, D.A., Z. Borghegyi, J. Csicsvari, A. Mamiya, K.D. 
Harris, and G. Buzsaki. 2000. Intracellular features predicted by extracellular
recordings in the hippocampus in vivo. J. Physiol. 84: 390-400.

\bibitem{Hines97} Hines, M.L., and N.T. Carnevale. 2000.  The NEURON simulation
environment.  Neural Computation 9: 1179-1209.

\bibitem{Hodgkin-Huxley} Hodgkin, A.L., and A.F. Huxley. 1952. A quantitative
description of membrane current and its application to conduction and excitation
in nerve.   J. Physiol. 117: 500-544.

\bibitem{Johnston97} Johnston, D., and S. Wu.  1997.  {\it Cellular
Neurophysiology}.  MIT Press, Cambridge MA.

\bibitem{Klee65} Klee, M.R., K. Offenloch, and J. Tigges. 1965. 
Cross-correlation analysis of electroencephalographic potentials and slow
membrane transients.  Science 147: 519-521.

\bibitem{Klee77} Klee, M., and W. Rall. 1977. Computed potentials of cortically
arranged populations of neurons.  J. Neurophysiol. 40: 647-666.

\bibitem{Koch99} Koch, C. 1999. {\it Biophysics of Computation}.  Oxford
University Press, Oxford UK.

\bibitem{Koch98} Koch, C. and I. Segev, editors. 1998. {\it Methods in Neuronal
Modeling} (2nd ed).  MIT Press, Cambridge MA.

\bibitem{Niedermeyer} Niedermeyer, E. and F. Lopes da Silva, editors. 1998. {\it
Electroencephalography} (4th ed).  Williams and Wilkins, Baltimore MD.

\bibitem{Nunez81} Nunez, P.L.  1981. {\it Electric Fields of the Brain.  The
Neurophysics of EEG}.  Oxford University Press, Oxford UK.


\bibitem{Peters91} Peters, A., S.L. Palay, and H.F. Webster.  1991.  {\it The
Fine Structure of the Nervous System}.  Oxford University Press, Oxford UK.

\bibitem{Numerical-Recipes} Press, W.H., H.P. Flannery, S.A. Teukolsky and W.T.
Vetterling. 1986. {\it Numerical Recipes. The Art of Scientific Computing}. 
Cambridge University Press, Cambridge UK.

\bibitem{Protopapas98} Protopapas, A.D., M. Vanier and J. Bower.  1998. 
Simulating large-scale networks of neurons.  In: {\it Methods in Neuronal
Modeling} (2nd ed), C. Koch and I. Segev, editors.  MIT Press, Cambridge MA. 
461-498.

\bibitem{Rall68} Rall, W. and G.M. Shepherd.  1968.  Theoretical reconstruction
of field potentials and dendrodendritic synaptic interactions in olfactory bulb.
J. Neurophysiol. 31: 884-915.

\bibitem{Ranck63} Ranck, J.B., Jr.  1963. Specific impedance of rabbit cerebral
cortex. Exp. Neurol. 7: 144-152.

\end{enumerate}
\end{document}